\pdfoutput=1
\documentclass[reprint,12pt,onecolumn,notitlepage,nofootinbib,floatfix]{revtex4-2}
\usepackage{amssymb}
\usepackage{amsmath}
\usepackage{amsfonts} 
\usepackage{bm}
\usepackage{color} 
\usepackage{graphicx}
\usepackage[bookmarks=false]{hyperref} 
\hypersetup{pdfstartview=FitH,pdfhighlight=/O,colorlinks=true}

\allowdisplaybreaks

\usepackage{epsfig} 
\usepackage{epstopdf}
\usepackage{latexsym}
\usepackage{psfrag}   
\usepackage{subfigure}  
\usepackage{booktabs}  
\usepackage{braket}
\usepackage{mathtools}
\usepackage{textcomp}
\usepackage{ifthen}
\usepackage{tensor}
\usepackage{protosem} 
\usepackage{wasysym}

\usepackage[toc]{appendix}
\usepackage{color,soul} 
\usepackage{datetime}

\definecolor{darkblue}{rgb}{0.2, 0, 0.8}
\definecolor{darkgreen}{rgb}{0.2, 0.71, 0}
\definecolor{awesome}{rgb}{1.0, 0.13, 0.32}
\definecolor{cadmiumred}{rgb}{0.89, 0.0, 0.13}
\definecolor{dukeblue}{rgb}{0.0, 0.0, 0.61}

\newcommand{\req}[1]{(\ref{#1})} 

\newcommand{\bea}{\begin{eqnarray}}
\newcommand{\eea}{\end{eqnarray}}
\newcommand{\ba}{\begin{eqnarray}}
\newcommand{\ea}{\end{eqnarray}}

\newcommand{\beq}{\begin{equation}}
\newcommand{\eeq}{\end{equation} }
\newcommand{\beqa}{\begin{eqnarray}}
\newcommand{\eeqa}{\end{eqnarray}}
\newcommand{\beqar}{\begin{eqnarray*}}
\newcommand{\eeqar}{\end{eqnarray*}}

\newcommand{\eg}{{\it e.g.,}\ }

\newcommand{\peff}{p_{\rm eff}}

\renewcommand{\href}[2]{#2}

\begin{document}

\title{Kasner Epochs, Eras and Eons}

\author{Pablo Bueno}
\email{Corresponding author: pablobueno@ub.edu}
\affiliation{Departament de F\'isica Qu\`antica i Astrof\'isica, Institut de Ci\`encies del Cosmos\\
 Universitat de Barcelona, Mart\'i i Franqu\`es 1, E-08028 Barcelona, Spain }

\author{Pablo A. Cano}
\email{pablo.cano@icc.ub.edu}
\affiliation{Departament de F\'isica Qu\`antica i Astrof\'isica, Institut de Ci\`encies del Cosmos\\
 Universitat de Barcelona, Mart\'i i Franqu\`es 1, E-08028 Barcelona, Spain }

\author{Robie A. Hennigar}
\email{robie.hennigar@icc.ub.edu}
\affiliation{Departament de F\'isica Qu\`antica i Astrof\'isica, Institut de Ci\`encies del Cosmos\\
 Universitat de Barcelona, Mart\'i i Franqu\`es 1, E-08028 Barcelona, Spain }

\date{\today}

\begin{abstract}  \vskip 0.2in 

As shown in the classic work of Belinski, Khalatnikov and Lifshitz, the approach to a generic space-like singularity in general relativity consists of a sequence of epochs and eras in which the metric is locally Kasner, connected by brief transitions. When quantum gravity effects are included, we argue that a new type of transition arises, giving rise to Kasner \emph{eons}: periods which are dominated by emergent physics at each energy scale. We comment on the different ways in which the Einsteinian eon may come to an end and show explicitly how additional Kasner eons arise in the interior of a black hole solution due to higher-derivative corrections.



\end{abstract}   

\maketitle 

\newpage

\section{Introduction}
Perhaps the most fundamental problem in theoretical physics is that of singularities. The classical description of gravity provided by general relativity breaks down at singularities necessitating a more complete physical theory to describe the universe in their vicinity. At the same time, singularities are ubiquitous. The Penrose-Hawking singularity theorems establish they generically arise in gravitational collapse and at the beginning of the universe, provided that certain global properties and energy conditions hold~\cite{Hawking:1973uf, Senovilla:1998oua}.

It is a widely held belief that singularities will be resolved by quantum effects. However, despite progress  in certain restricted circumstances, see {\it e.g.}~\cite{Ashtekar:2022oyq, Bambi:2023try}, there is no known mechanism that justifies this belief on general grounds. In fact, very little is known about quantum effects in the vicinity of generic singularities, besides the obvious point that they will be important. 


A potential avenue for progress consists in understanding the implications of quantum effects on universal aspects of gravity near singularities. That anything \textit{universal} can be said about the behaviour of \textit{generic} singularities is not at all obvious. Nonetheless, more than 50 years ago, Belinski, Khalatnikov, and Lifshitz (BKL) gave a complete, local description of gravitational dynamics near a space-like singularity~\cite{Belinsky:1970ew}. Their analysis revealed two remarkable and universal features: ultra-local and oscillatory, chaotic dynamics. These originally quite speculative ideas now have considerable support coming largely from numerical work~\cite{Berger:1993ff, Weaver:1997bv,Berger:1998vxa, Berger:2002st, Garfinkle:2003bb,  Garfinkle:2020lhb} but also more rigorous results in certain limiting cases~\cite{Cornish:1996hx, Rendall:1997dc, Andersson:2000cv, Damour:2002tc}.

%

The dynamics becomes ultra-local in the sense that each spatial point evolves independently of the others, which can be heuristically thought of as the veritable {\it shredding} of space-time. This is often phrased as time derivatives dominating spatial derivatives in the equations of motion, or equivalently as an emergent Carrollian regime.


Key to understanding the evolution of the oscillatory metric is the (generalized) Kasner geometry~\cite{Lifshitz:1963ps}. The Kasner solution describes a homogeneous but anisotropic cosmology that expands in one spatial direction while contracting in all other spatial directions. In the case of the generalized Kasner geometry, the exponents governing the expansion/contraction are allowed to depend on space.  The metric takes the form
\begin{equation}\label{Kas}
\mathrm{d}s^2=-\mathrm{d}\tau^2+\sum_{i=1}^{D-1}a_{i}(\tau)^2 (e^{i})^2\, ,
\end{equation}
where $e^{i}$ are frame vectors, and because of ultralocality, each of the scale factors $a_{i}(\tau)$ can evolve independently at each spatial point. Whenever the spatial curvature is negligible, the metric becomes Kasner, with $a_{i}(\tau)=\tau^{p_{i}(x)}$ and where the Kasner exponents satisfy
\begin{equation}\label{KasnerEinstein}
\sum_{i=1}^{D-1}p_{i}=\sum_{i=1}^{D-1}p_{i}^2=1\, .
\end{equation}

The full evolution consists of an infinite sequence of Kasner epochs and eras connected by brief transitions in which the Kasner exponents change --- but always satisfy \eqref{KasnerEinstein}~\cite{Belinski:2017fas}. An epoch is a period of time during which the space-time metric is approximately given by a Kasner geometry. Epochs generically come to an end when growing perturbations due to spatial curvature become important, driving the universe from one Kasner epoch to another. During the transition between epochs, the Kasner exponents shift in a known way, with the direction of expansion switching with one of the directions of contraction. Kasner eras emerge over a longer time scale. An era contains potentially a large number of epochs. The defining characteristic is that during an era, all epochs swap expansion/contraction  between the same two directions, repeatedly. In the four-dimensional case, the equations governing the dynamics are equivalent to those for the Bianchi-IX --- or Mixmaster --- universe~\cite{Misner:1969hg}.

Of the two features revealed by the BKL analysis, ultra-locality is more fundamental. The chaotic nature of the singularity can be influenced by certain matter content and space-time dimension~\cite{Demaret:1985jnc}. However, the quantum fate of both properties are natural --- and important --- questions. 

One ubiquitous manifestation of quantum effects is higher-derivative corrections to the classical equations of motion \cite{tHooft:1974toh,Goroff:1985th,Gross:1986mw,Grisaru:1986vi,Sakharov:1967nyk,Visser:2002ew,Endlich:2017tqa}. These corrections will lead to modifications of the BKL analysis as it applies to general relativity. To see this is necessarily the case, consider corrections to the action in a derivative expansion, as in effective field theory. On a Kasner background, terms with $n$-derivatives will contribute to the equations of motion as $\tau^{-n}$, where $\tau$ is the proper-time in the Kasner metric. Since the powers of $\tau$ are unequal for different values of $n$, the higher-derivative terms will become important, ultimately overwhelming the two-derivative sector.
This leads to the end of the Einstein gravity \emph{eon} --- the period of time where evolution is driven by Einstein equations.

If higher-derivatives lead to a break-down of the BKL picture, then exactly how does that occur? What replaces it? Laying out possible answers to these questions and providing explicit examples is the purpose of this Essay.

\section{ Terminating BKL: three scenarios}\label{sec2}
The classical Einsteinian description of spacetime in the proximity of a singularity is expected to receive modifications of various kinds. On the one hand, quantum fluctuations of the metric should become relevant at the Planck scale, eventually leading to a complete breakdown of the classical spacetime description and, hypothetically, to a full resolution of the singularity. Notwithstanding this, it is plausible that there exists an intermediate regime in which the correct description is provided by a classical metric which evolves following a set of modified Einstein equations. Such corrections would be associated to intermediate energy scales and would take the form of higher-curvature terms in the gravitational action --- prototypically, this is what happens with  $\alpha^{\prime}$ corrections in string theory. 

A logical possibility is that the fully quantum/stringy regime is reached without going through any such additional phases. Here we would like to explore the alternative situation, namely, the effects that an intermediate higher-curvature-dominated phase may have on the fate of spacetime near a singularity.

From this perspective, we will outline three ways by which the Einsteinian BKL eon may come to an end. The list is not exhaustive, but it is motivated by the fact that these possibilities arise already in the simplest possible circumstances but are not consequences of those circumstances.

{\bf Kasner eons.}  When the comoving volume of spatial slices becomes smaller than a certain threshold, the effects of higher-derivative terms kick in, which follows from the fact that such terms scale as inverse powers of the volume. This typically takes place at a proper time $\tau\sim \ell$ away from the singularity, where $\ell$ is the length scale of new physics.  Assuming one can still make sense of a classical evolution beyond this time, one would find that the relations for the Kasner exponents \req{KasnerEinstein} will suffer a sharp transition to a new set of constraints. This marks the transition from the Einstein gravity eon to a higher-derivative eon, where the dynamical evolution is drastically altered. In this new eon we also expect to have an analogous BKL picture: an ultra-local singularity possibly with Kasner eras and epochs governing the approach, but with modified Kasner transition rules.  

Now, let us imagine that there are subsequent corrections that appear at even shorter length scales $\ell_{1}$, $\ell_{2}$, \ldots, so that we have a hierarchy of scales $\ell\gg\ell_{1}\gg\ell_{2}\ldots$ Then, we will have a cascade of transitions between different Kasner eons --- each possibly with their particular sets of exponents, eras and epochs --- which will happen roughly at $\tau_{i}\sim \ell_{i}$ before the singularity. On the other hand, it may happen --- and it is probably the most natural possibilty --- that all these corrections become relevant at the same scale $\ell$. In that case one can interpret the time evolution as a continuous transition between all these Kasner eons, leading to an effective time dependence of the Kasner exponents. 

In general, we expect that the existence of this sequence of Kasner eons pushes the singularity forward in the future. In particular cases, with an infinite sequence of eons, one could even hope that the singularity is pushed infinitely far away --- and hence, resolved. This would be analogous to time-reversal of~\cite{Arciniega:2018tnn}.

{\bf Finite-volume singularity.} The previous scenario assumes that the universe keeps contracting down to arbitrarily small size and it reaches a singularity when the volume of spatial slices vanishes. However, one may conceive of a scenario in which the singularity appears at a finite volume. This will happen if the metric remains finite but the curvature diverges, hence leading to diverging tidal forces. 

Singularities of this type were first identified in higher-curvature theories of gravity in~\cite{Kitaura:1990ve, Kitaura:1993cm}, where they were shown to  occur within the Bianchi-I family of metrics. Therefore it is quite plausible that such singularities could play a role in terminating the BKL scenario, though there currently exists no construction of \textit{generic} singularities of this form.

{\bf Inner horizons.} A third possibility is that the introduction of higher-curvature terms qualitatively changes the character of the near-singularity geometry, replacing the space-like singularity with a Cauchy horizon and possibly subsequent time-like singularity. The Cauchy horizon would be prone to the same type of mass inflation instability that occurs in General Relativity~\cite{Poisson:1990eh}, likely resulting in a null singularity. Behaviour of this type is certainly plausible in a higher-curvature theory, as it can happen already in Einstein gravity~\cite{Dafermos:2012np} and plays an important role understanding the interior of the Kerr black hole~\cite{Dafermos:2017dbw}. 

Null singularities are also generic in Einstein gravity~\cite{Ori:1995nj, Luk:2013cqa}. It is therefore conceivable that the Einsteinian eon, after transitioning to a higher-derivative eon, ultimately ends at a null curvature singularity. Nonetheless, the complete story is less clear in this case. For example, there exist explicit constructions wherein a would-be null singularity is replaced by a space-like singularity due to classical~\cite{Hamilton:2017qls} or quantum effects~\cite{Emparan:2021yon}. In any case, it would be worthwhile to better understand the local gravitational dynamics in the vicinity of null curvature singularities, both in general relativity and beyond --- see~\cite{Ori:1999phc}. 


\section{A toy model}



We expect the above scenarios will take place for generic space-like singularities, but many open problems must be addressed before this can be confirmed.\footnote{The task of building explicit models exhibiting both near-singularity BKL-like behavior and an event horizon in the same solution has proven to be notably challenging --- see \cite{DeClerck:2023fax} for a recent exception.} Nonetheless, here we show that these features appear already under the simplest possible circumstance of a spherically symmetric black hole interior. The idea is then to study the effect of higher-curvature corrections on the 
near-singularity behavior of the Schwarzschild solution.  A natural setup for doing this is that of Lovelock gravities, which are the most general diffeomorphism invariant theories of gravity which possess second-order equations for the metric \cite{Lovelock:1971yv,Lovelock:1972vz}. Amongst these, the simplest non-trivial scenario occurs in five dimensions, where the Lovelock action reads
\begin{equation}
S=\frac{1}{16\pi G} \int \mathrm{d}^5x \sqrt{|g|} \left(R + \lambda \mathcal{X}_4  \right)\, , \quad \text{where} \quad \mathcal{X}_4\equiv R^2 -4 R_{ab}R^{ab}+R_{abcd}R^{abcd}\, ,
\end{equation}
is the Gauss-Bonnet (GB) density,  $\lambda$ is a coupling constant with dimensions of length$^2$ and $G$ is the Newton constant, which in five dimensions has dimensions of length$^3$.

\begin{figure}[t!] \hspace{-0.7cm}
\includegraphics[scale=0.78]{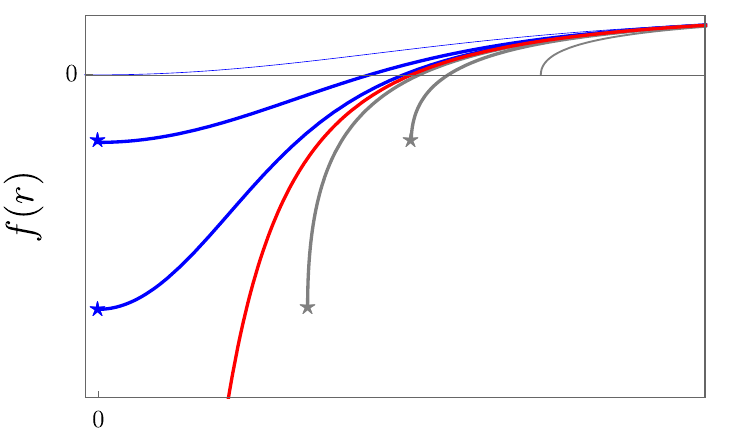}
\caption{ Metric factor $f(r)$ for Einstein-Gauss-Bonnet black holes. Blue and gray curves represent solutions with $\lambda>0$ and $\lambda <0$, respectively. The red curve corresponds to the Schwarzschild solution. The stars represent curvature singularities. The thin lines correspond to the limiting cases for which $|\lambda| = 4GM/(3\pi)$. }
\label{fig1}
\end{figure}

 This theory admits a static and spherically symmetric generalization of the five-dimensional Schwarzschild black hole whose interior can be described by the metric
 \begin{align} \label{intf}
\mathrm{d}s^2 =&\, \frac{\mathrm{d} r^2}{f(r)} -f(r) \mathrm{d} z^2 + r^2 \mathrm{d}  \Omega_{\mathbb{S}^{3}}^2 \, ,
\end{align}
where \cite{Boulware:1985wk,Wheeler:1985nh},
\begin{equation}
f(r)= 1 + \frac{r^2}{4 \lambda} \left[1-\sqrt{1+ \frac{64 G M\lambda}{3\pi r^4}} \right] \, .
\end{equation}
The structure of the solution is determined by two characteristic values of the radial coordinate, namely,
\begin{equation}
r_{\rm h}\equiv \left(\frac{8 GM}{3\pi} -  2 \lambda\right)^{1/2} \, ,\qquad   r_{\star} \equiv \left( \frac{64 GM |\lambda|}{3\pi}\right)^{1/4}\, .
\end{equation}
Depending on the magnitude and sign of $\lambda$, and assuming a positive mass $M$, one finds the following situations --- see Fig.\ref{fig1}. Whenever $|\lambda| > 4GM/(3\pi)$, the solution describes a naked singularity at $r_\star$. On the other hand, when $0\leq  \lambda< 4GM/(3\pi)$, there is an event horizon at $r_{\rm h}$ which hides a curvature singularity at $r=0$. Finally, when $-4GM/(3\pi)< \lambda <0$, the solution has an event horizon at $r_{\rm h}$ which hides a curvature singularity at $r_{\star}$. We restrict our analysis to the cases in which there is an event horizon.

Performing the change of variables $-\mathrm{d}\tau^2= \mathrm{d}r^2/f(r)$ in the black hole interior, it is easy to see that whenever $f(r) \sim r^{-s}$,  the spacetime is effectively described by a Kasner metric with exponents $p_1=p_{\rm eff}$, $p_2=p_3=p_4=p_{\rm eff}+1$, where we define the  ``effective'' Kasner exponent --- which corresponds to the $\mathrm{d}z^2$ component of the metric --- as
\begin{align}
\peff(r) \equiv \frac{r f'(r)}{2 f(r) - r f'(r)} \, .
\end{align}
Below we use this quantity to probe the presence of eons in the solution. These will be manifest in periods of approximately constant values of  $\peff$ as we approach the singularity.

\begin{figure}[t!] \hspace{-0.7cm}
\includegraphics[scale=0.85]{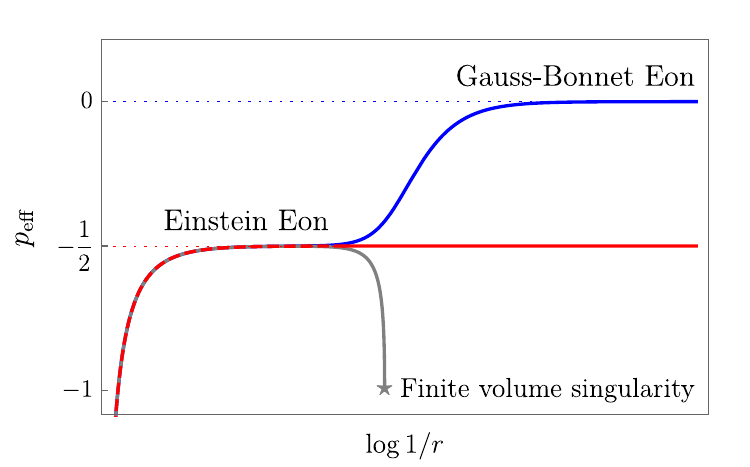}
\caption{ Effective Kasner exponent as a function of $\log(1/r)$ for Einstein-Gauss-Bonnet black hole interiors. In the absence of the higher-curvature correction, the near-singularity solution tends to the Einstein gravity Kasner metric with $p_1=-1/2$. In the presence of $\lambda$, the Einsteinian behavior dominates the dynamics during an eon which is terminated  at $r_{\rm end} \lesssim r_{\star}$. For $\lambda >0$, the metric transitions to a new Gauss-Bonnet-dominated eon which approaches a Kasner solution of the pure Gauss-Bonnet theory  with $p_1=0$. On the other hand, for $\lambda <0$, the Einsteinian eon is terminated by a spacetime finite-volume curvature singularity at $r= r_{\star}$.    }
\label{fig2}
\end{figure}

Whenever $\lambda$ is positive, the singularity remains at the coordinate point $r=0$. However, the proper time it takes an infalling observer to reach it increases with the magnitude of $\lambda$. As we approach the singularity, there is a first phase in which the Einstein gravity contribution dominates, eventually undergoing an eon in which it behaves as a Kasner metric with $p_1=-1/2$, $p_2=p_3=p_4=1/2$ --- namely, as a pure Kasner solution to Einstein gravity. As we move even closer, this phase comes to an end roughly at $r_{\rm end} \sim  r_{\star}$. From that point on, Gauss-Bonnet dynamics takes over, and spacetime undergoes a new eon in which it behaves as a Kasner metric with exponents $p_1=0$, $p_2=p_3=p_4=1$. These are precisely the values of the exponents of a  Kasner solution to five-dimensional GB gravity (without Einstein-Hilbert term)~\cite{Camanho:2015yqa}.
In the absence of the correction, the Einstein eon would last all the way till the singularity --- see Fig.\,\ref{fig2}. The singularity itself  gets softened with respect to Einstein gravity. For instance, the Kretschmann invariant diverges as $R_{abcd}R^{abcd}\sim 1024 (GM)^2/(3\pi^2 r_{\star}^4 r^4)$ as opposed to the Einsteinian behavior, $R_{abcd}R^{abcd}\sim 512 (GM)^2/(\pi^2 r^8)$. 


 Observe that while the GB eon always exists as long as $\lambda >0$, the existence of the Einsteinian one relies on  $\lambda$ being much smaller than $GM$. Otherwise, $p_{\rm eff}$ will transition directly to the GB eon without an intermediate Einstein phase. A qualitatively analogous structure of eons is found for higher-dimensional Lovelock theories \cite{KasnerLove}. In particular, including higher-curvature terms gives rise, for certain values of the couplings, to additional eons characterised by effective exponents which coincide with the corresponding Kasner solutions of each Lovelock density. This leads to a ``stair-like'' structure for $p_{\rm eff}$ similar to the one displayed in Fig.\,\ref{fig2} but with additional stairs corresponding to each of the higher-curvature densities.  Depending on the relative magnitudes of the couplings, some of the eons may be skipped.  As more terms are added to the action, the proper time to the singularity keeps on increasing. In the limit of infinitely many terms, it would take infinite proper time to reach the singularity, effectively resolving it. 

On the other hand, for negative values of $\lambda$ the spacetime terminates at $r_\star$, which is the point beyond which the metric function $f(r)$ would become complex. This is an instance of the second scenario presented in Section \ref{sec2}. The Einstein gravity eon is now terminated before than naively expected at a point of finite spatial volume and without going through any additional eon.  At that point there is a curvature singularity which is softer than the one corresponding to positive values of $\lambda$.  Indeed, one finds $R_{abcd}R^{abcd}\sim 64 (GM)^2/[9\pi^2 r_{\star}^5 (r-r_{\star})^3]$ near $r=r_\star$.  In this case, the proper time to the singularity gets reduced with respect to the naive Einstein gravity one. 

The last scenario proposed in Section \ref{sec2} cannot be achieved within the present model, as the introduction of $\lambda$ never gives rise to an additional horizon. In $D\geq 7$, the introduction of the cubic Lovelock density in the action gives rise to a modified solution which does contain an inner horizon for certain values of the higher-curvature couplings --- see \eg \cite{Camanho:2011rj}.




\section{Discussion}
We have introduced the notion of Kasner eon as the period during which --- in the vicinity of a space-like singularity --- spacetime is described by a classical Kasner-like geometry like (\ref{Kas}) with $a_i(\tau)=\tau^{p_i(x)}$ and where the Kasner exponents $p_i(x)$ are determined by the dynamics of certain higher-curvature correction to Einstein gravity. As the singularity is approached, the Einsteinian eon --- during which the BKL description  holds --- will be terminated as a result of the corrections becoming relevant. 

We have explored three possible scenarios for the  phase arising after the termination of the Einsteinian eon: a (possibly never-ending) tower of higher-curvature Kasner eons characterized by modified Kasner exponents and transition rules; the emergence of a finite-volume curvature  singularity;  and the appearance of inner horizons which may ultimately result in a null singularity. This list is not intended to be completely exhaustive. Additional possibilities --- such as bounces --- may also occur. However, here we have chosen to focus on those possibilities that are manifest already in the simplest cases. 

Using effective field theory constraints it may be possible to probe which of these is more plausible in realistic scenarios. In particular, studying the sign of the leading correction to the Einstein-Hilbert action may be able to discriminate  between them. Indeed, it is natural to speculate that whenever the effective Kasner exponent is not monotonically increasing as the singularity is approached, spacetime terminates at a finite-volume singularity. If this conjecture holds true, the behavior of the effective Kasner exponent at the end of the Einstenian eon would tip the scales in favor of one of the possibilities. Alternatively, since it is now known that Lovelock theories admit a well-posed initial value problem~\cite{Kovacs:2020ywu}, one could revisit the classic numerical studies of the BKL phenomenon now supplemented by higher-curvature corrections, \textit{e.g.}, a Gauss-Bonnet term. This could allow for a direct exploration of the end of the Einsteinian eon, as well as the implications of non-local phenomena such as spikes.

In the toy model presented here, we have considered the interior of a static and spherically symmetric black hole. This is blind both to the ultralocal nature of spacetime as well as to the chaotic BKL behavior  insofar as the effective Kasner exponents are identical at every spatial point for fixed $\tau$ and there is a single stable Kasner epoch per eon. Understanding how these features get modified for higher-curvature Kasner eons would require a detailed study of less symmetric near-singularity solutions.  This would provide additional hints on the ultimate fate of the Einsteinian eon.

%

%

\vskip 0.2in
\emph{Acknowledgments:}
The authors are thankful to Roberto Emparan and David Mateos for useful conversations related to the topics of the essay. RAH is also grateful to Sean Hartnoll and Harvey Reall.  PB was supported by a Ram\'on y Cajal fellowship (RYC2020-028756-I) from Spain's Ministry of Science and Innovation. The work of PAC received the support of a fellowship from “la Caixa” Foundation (ID 100010434) with code LCF/BQ/PI23/11970032. 
The work of RAH received the support of a fellowship from ``la Caixa” Foundation (ID 100010434) and from the European Union’s Horizon 2020 research and innovation programme under the Marie Skłodowska-Curie grant agreement No 847648 under fellowship code LCF/BQ/PI21/11830027. Research supported by grant PID2022-136224NB-C22, funded by MCIN/AEI/ 10.13039/501100011033/FEDER, UE.


\renewcommand{\leftmark}{\MakeUppercase{Bibliography}}
\phantomsection
\bibliographystyle{JHEP}
\bibliography{kasner}
\label{biblio}

\end{document}